# The Trigger System of the CMS Experiment

Marta Felcini on behalf of the CMS Collaboration

**Abstract**

We give an overview of the main features of the CMS trigger and data acquisition (DAQ) system. Then, we illustrate the strategies and trigger configurations (trigger tables) developed for the detector calibration and physics program of the CMS experiment, at start-up of LHC operations, as well as their possible evolution with increasing luminosity. Finally, we discuss the expected CPU time performance of the trigger algorithms and the CPU requirements for the event filter farm at start-up.

# The Trigger System of the CMS Experiment

Marta Felcini*

*University College Dublin, Dublin, Ireland*

on behalf of the CMS Collaboration

**Abstract**

We give an overview of the main features of the CMS trigger and data acquisition (DAQ) system. Then, we illustrate the strategies and trigger configurations (trigger tables) developed for the detector calibration and physics program of the CMS experiment, at start-up of LHC operations, as well as their possible evolution with increasing luminosity. Finally, we discuss the expected CPU time performance of the trigger algorithms and the CPU requirements for the event filter farm at start-up.

## 1. Introduction

The CMS detector [1] is now built and in its final commissioning phase [2], preparing to collect data from the proton-proton collisions to be delivered by the Large Hadron Collider (LHC), at a centre-of-mass energy of up to 14 TeV. The CMS experiment employs a general-purpose detector with nearly complete solid-angle coverage, which can efficiently and precisely measure electrons, photons, muons, jets (including tau- and b-jets) and missing energy over a wide range of particle energies and event topologies. These characteristics ensure the capability of CMS to cover a broad programme of precise measurements of Standard Model physics and discoveries of new physics phenomena. The trigger and data acquisition system must ensure high data recording efficiency for a vast variety of physics objects and event topologies, while applying online very selective requirements.

The CMS trigger and data acquisition system [3,4]  is designed to cope with unprecedented luminosities and interaction rates. At the LHC design luminosity of $10^{34}$cm$^{-2}$s$^{-1}$ and bunch-crossing rates up to 40 MHz, an average of about 20 interactions will take place at each bunch crossing. The trigger system must reduce the bunch crossing rate to a final output rate of O(100) Hz, consistent with an archival storage capability of O(100) MB/s.

The trigger configurations (trigger selection algorithms and their parameter settings) must be chosen and optimized to address the detector needs and physics objectives of the experiment, depending on luminosity, machine and detector conditions. According to the LHC start-up plan, the LHC instantaneous luminosity (hereafter referred to as luminosity L), in the initial phase, is expected to increase gradually before reaching the design luminosity. Runs at low luminosities will be useful to fully commission and calibrate the detector as well as to measure Standard Model processes, before reaching the high luminosity phase, when discoveries of new physics phenomena will be the main goal of the experiment.

In this article, after a concise description of the CMS trigger and data acquisition (DAQ) system, we discuss the strategies and trigger configurations (trigger tables) developed for the CMS detector calibration and physics program, at start-up, as well as their possible evolution with increasing luminosity. We also discuss the expected CPU time performance of the trigger algorithms and the requirements for the events filter farm at start-up.

---

* Corresponding author. Tel.:+41 22 7670641; fax: +41 22 7669191.
*E-mail address*: marta.felcini@cern.ch.

## 2. The Trigger and DAQ system

The CMS trigger architecture employs only two trigger levels (not three or more levels as in more traditional systems [5]). The Level-1 Trigger (L1T) [3] is implemented using custom electronics. The High Level Trigger (HLT) [4] is implemented on a large cluster of commercial processors (the Event Filter, EF, farm).

The L1T system must process information from the CMS detector at the full bunch crossing rate (up to 40 MHz at the highest LHC luminosities). The time between two successive bunch crossings, along with the wide geographical distribution of the electronic signals from the CMS sub-detectors, require the use of fast electronics. The time for processing the detector information in the L1T system is limited by the front-end (FE) electronics capability to store the detector data during the L1T decision process. The FE electronics modules can store the data from up to 128 contiguous bunch crossings, i.e. ~3 μs. Within this time interval, the detector information must be transferred to the L1T processing elements, the decision must be formed and the decision signal must be transferred back to the FE electronics. The resulting time available for processing the data in the L1T system is no more than ~1μs. Thus the L1T can process a limited amount of detector data, from calorimeters and muon chambers, with coarser granularity and lower resolution than the full information recorded in the FE electronics. The processing elements of the L1T system are custom-designed. Details of the architecture, the design and the selection algorithms in the L1T can be found in [3]. The L1T system is designed to achieve a bunch crossing rate reduction factor of up to 400, for a maximum mean event accept rate of 100 kHz. The estimated average size of an event record is O(1MB). After the acceptance of an event by the L1T, about 700 FE modules store the event data, each carrying 1-2 kB of data per L1T accepted event.

The next online selection step, the HLT [4], must operate a rate reduction of 1000, dictated by the ability to store and reconstruct data offline at a maximum accept rate of O(100) Hz, or O(100) MB/s. Such a rejection factor requires that the HLT selection be based on full granularity and resolution information from the whole detector, including trackers, with selection algorithms almost as sophisticated as those used in the offline event reconstruction. This implies the usage of fully programmable commercial processors for the execution of the HLT. The expectation that the HLT algorithms will demand a mean processing time of O(10) ms, along with the maximum HLT input rate of 100 kHz, implies that O(1000) processors in the EF farm must be employed for this processing stage. This, in turn, dictates that the Data Acquisition (DAQ) system [4] must provide the means to feed data from ~700 FE modules to about 1000 processors, at a sustained bandwidth of up to 100 kHz×1MB=100 GB/s. The interconnection of such a large number of elements, at such a bandwidth, implies the usage of a switching network (Builder Network). Two systems complement this flow of data from the FE memories to the EF farm: the Event Manager, responsible for the actual data flow through the DAQ, and the Control and Monitor System, responsible for the configuration, control and monitor of all the elements. The architecture of the CMS Trigger and DAQ system is shown schematically in **Figure 1.**

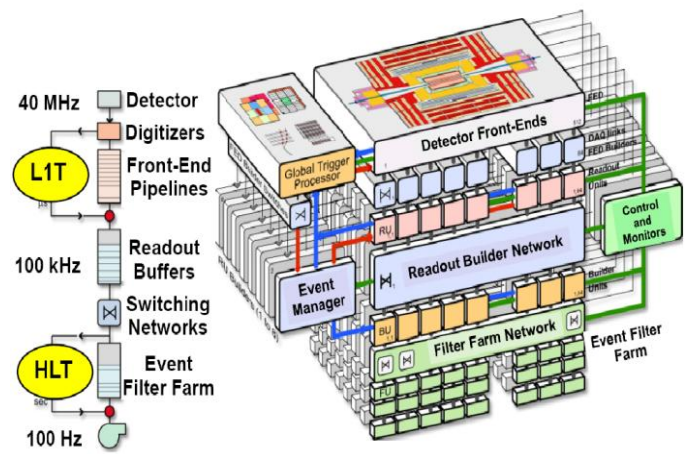

**Figure 1**. Schematic view of the CMS trigger and DAQ system showing, left, the successive stages and, right, the modularity (slices) of the system [4].

At the start-up of the LHC, the CMS DAQ system is expected to be able to sustain an event readout rate of up to 50 kHz from the L1T. Events processed by the EF farm, running HLT reconstruction and selection algorithms, will be accepted at a rate of up to 300 Hz for output to permanent storage. In the following, we report the results of a detailed study **[6]** about the expected physics and computing performance of the HLT selection algorithms at LHC startup luminosities of $O(10^{32})$ cm$^{-2}$s$^{-1}$.

## 3. Trigger Criteria and Trigger Performance

The trigger decision ("accept" or "reject") is based on the characteristics of the trigger objects (candidate muons, electrons, photons, jets, etc.) identified and measured using the detector information available at the trigger level. Coarse detector information, from calorimeters and muon chambers, is available at the L1T. At the HLT, the complete information from the whole detector is available to be processed by off-line quality algorithms. The trigger selections are implemented as trigger "paths". If the event passes one or more of these paths, it is accepted for permanent storage. A trigger path is a set of algorithms which reconstruct one or more candidate objects and apply selection criteria to the reconstructed quantities. A trigger path is constituted by two types of modules: producer and filter modules. Producer modules produce, or reconstruct, the trigger primitives (quantities used for the trigger decision). Filter modules apply selection criteria to the HLT reconstructed quantities. There is a third type of module, called prescaler, whose action is to apply a determined prescale factor, so that when a trigger path is prescaled by a factor of N, only one out of N events is considered for processing by that trigger path. A prescale factor is applied to a trigger path to reduce its output rate and keep the overall rate within the allocated bandwidth budget. A set of trigger paths, in a defined configuration, constitutes a trigger table, which is characterized by its paths, the values of the thresholds used in the filters and the prescale values.

Trigger criteria (thresholds and prescale values) must be chosen to maximize the trigger capabilities depending on luminosity, collider and detector conditions. The trigger performance is measured by three quantities: (a) the "background" rate, which must be kept low; (b) the "signal" efficiency, which must be kept high; (c) the CPU time consumption by the trigger algorithms, which must be kept low, to avoid dead-time and inefficiencies. The definitions of "background" and "signal" change depending on the goals of the experiment, at different luminosities, as shown schematically in **Table 1**. At startup, when luminosities are expected to be relatively low, the wanted "signal" is constituted mainly of events from SM (QCD, heavy flavor and W/Z production) processes, needed for detector calibration, as well as for measurements of the known physics processes, at a centre-of-mass energy never attained before. As the luminosity increases, the bulk of the SM processes will be considered "background", to be reduced, in favor of more exotic "signal"-like events (e.g. events with high transverse momentum, $P_T$, objects and/or high object multiplicity), as signatures of expected, or unexpected, new physics phenomena.

The actual trigger performance will be measured and optimized with real data from collisions, when the actual experimental (collider and detector) conditions will be known. In preparation for data taking, we can use our present best knowledge of the detector response and possible collider condition scenarios to study and optimize trigger criteria, in view of adjusting them when real collisions and detector data will be available. The flexibility of the trigger system allows to introduce modifications in an efficient manner for optimal performance adapted to the actual running conditions while taking data. The robustness of the algorithms which determine the trigger primitives also ensure that the system will not be too sensitive to detailed changes with respect to the expected conditions.

## 4. Development of trigger tables for early physics

**Table 2** gives an overview of the "ingredients" composing a trigger table. Different types of triggers can be used to compose the trigger table. We can broadly classify the triggers according to the type and the number of objects used for the trigger decision: (1) single-object triggers, like single-jet, single-muon, single-electron, (2) double (or multiple)-object triggers, using two (or more) objects of the same type, like two-electron, or three-muon, or four-jet triggers, and (3) cross-object triggers, which may use any combination of objects of different types, like electron-plus-three-jets or muon-plus-tau-jet triggers. A trigger type may identify a single path or a set of trigger paths. The example shown in the second column of **Table 2** represents a set of three non isolated single muon (NoIsoMuon) trigger paths. Each of them requires a reconstructed non isolated single muon (i.e. it uses the same muon producer) but applies a different threshold on the muon $P_T$, e.g. from a lowest value of P3, to an intermediate value of P2, to the highest threshold value of P1 (if required, more than three paths may be included in a set). The lowest $P_T$ triggers may be left unprescaled, if their rate is acceptably low, at low luminosities. As the luminosity increases, they will be prescaled, while leaving unprescaled only the highest $P_T$ muon trigger path. The advantages of such a configuration, with sets of paths, using all the same trigger object(s), but each applying different threshold value(s), are flexibility and manageability of the trigger table, as well as optimal use of the available bandwidth at any luminosity. Indeed the trigger table configuration, i.e. the sets of trigger paths and their thresholds, can be kept the same over a large range of luminosities. The only parameters which need to be changed, as the luminosity changes, are the prescale factors. Such a configuration also allows the application of dynamic prescale factors, during a collider run, where luminosity can change significantly.

| Luminosity range | Goals of the experiment | Definition of Signal | Definition of Background |
|---|---|---|---|
| from ~$10^{29}$s$^{-1}$cm$^{-2}$ to ~$10^{31}$s$^{-1}$cm$^{-2}$ | Calibrate and align all the sub-detectors | QCD events | Machine, detector noise |
| from ~$10^{31}$s$^{-1}$cm$^{-2}$ to ~$10^{32}$s$^{-1}$cm$^{-2}$ | Measure SM physics: b-physics, W/Z, top. Use SM processes for more refined detector calibration. "Early" discoveries. | SM processes: b-physics, W/Z, top production. "Early" (high cross-section) new physics. | As above and QCD events |
| from ~$10^{32}$s$^{-1}$cm$^{-2}$ to ~$10^{34}$s$^{-1}$cm$^{-2}$ | Detector fully calibrated and detector response very well understood. Full exploration of the high energy frontier and new physics discoveries. | Expected new physics: Higgs, Supersymmetry, Extra dimensions… Unknown new physics: ??? | As above and SM processes |

**Table 1**: Overview of the goals of the experiment and the corresponding definition of signals and background, depending on the luminosity range. Based on these definitions, the trigger performance is evaluated.

| Trigger type /stream | Trigger conditions | | Purpose |
|---|---|---|---|
| **Inclusive** **Single object** IsoMuon NoIsoMuon RelaxElectron Electron Others…. | $P_T$ Threshold | Prescale factor | High $P_T$ triggers used for: high-$P_T$ signals (W,Z, top, Higgs, … ) Low $P_T$ triggers used for: - Low $P_T$ signals (b-physics…) - Background measurements - Trigger efficiency measurement - Detector calibration |
| | P1 | 1 | |
| | P2 (< P1) | N2 (>1) | |
| | P3 (<P2) | N3 (>N2) | |
| **Double-object** DoubleNoIsoMuon DoubleRelaxElectron Others…. | Rate vs Pt plot: P3, N3 ; P2, N2 ; P1, 1 | | |
| **Exclusive** **Cross-object** NoIsoMuon+3jets RelaxElectron+NoIsoMuon Others…. | Low $P_T$ thresholds allowed (lower than for Inclusive single object triggers) because of low rate. Prescale factors may be needed at higher luminosities | | Cross-object triggers used for specific selections of interest both for Standard Model process and for new physics searches |

**Table 2**: Overview of the different trigger types which compose a trigger table, the trigger conditions (thresholds and prescales), and their purpose. Each trigger type produce an associated trigger data stream.

Each trigger type, e.g. single muon, determines the corresponding trigger data stream, i.e. the data set including all the events passing one or more of the trigger paths of that type. The fact that the trigger table performances is adjusted to luminosity only by modification of the trigger prescales, and not the trigger thresholds, is also a significant advantage for the off-line data analysis.

Single- and double-object trigger paths, especially at low luminosities, are set up to be as inclusive as possible. Loose trigger criteria will be used to collect data over a broad range for trigger, detector and physics studies. Cross-object triggers are dedicated, exclusive triggers, generally designed to select specific topologies. Because of the multiple object requirements, the accept rates of these triggers are expected to be relatively small, as compared to single object triggers. Thus, comparatively lower thresholds can be used, at small bandwidth cost.

We design trigger tables (configurations of trigger paths, thresholds and prescales), to optimize the trigger performance, depending on the expected luminosity and expected pile-up conditions, as well as on the expected detailed response of all sub-detectors. A trigger table must be designed for a maximum trigger accept rate. The allowed bandwidth has to be shared among the different triggers according to detector and physics priorities at a given luminosity (see **Table 1**). The general procedure to design a trigger table is outlined here: (a) we start with fully simulated events of all known physics processes (QCD, W/Z, top production), including the effect of pile-up (overlapping events within one bunch-crossing), which depends on luminosity; (b) for each event, we simulate the actions of the L1T and HLT, reconstructing candidate objects (electrons, muons, jets, etc.) coarsely at the L1T and precisely at the HLT, and applying the trigger criteria at each trigger level; (c) for given luminosity and pile-up conditions, we calculate trigger rates for all (single object, double object, multiple object, cross object) triggers as a function of trigger thresholds applied at each trigger level; (d) depending on the goals of the experiment at a given luminosity, we allocate the bandwidth sharing among the different triggers; (e) the bandwidth allocation defines thresholds and prescales for the different trigger paths in the table.

### 4.1 L1T Rates and Tables

L1T tables are designed for a maximum L1 accept rate of 17 kHz, which is 1/3 of the initial DAQ readout capability of 50 kHz from the L1T. This safety factor is introduced to keep into account the uncertainty on the predicted rates and other unknown factors. **Figure 2** shows the expected L1T rates at $L=10^{32}$ cm$^{-2}$s$^{-1}$ for L1T single-object triggers as a function of the trigger object transverse momentum threshold. We observe that: (a) muon trigger rates are the lowest, down to very small muon $P_T$ threshold values; (b) electron rates are expected to be larger, especially at low electron transverse energy, $E_T$, threshold values; (c) jet rates are high over the whole jet $E_T$ range. For double-object triggers (not shown in **Figure 2**), as well as for cross object triggers, rates are one to more orders of magnitude lower than for the corresponding single object triggers [6]. Thus, the thresholds of double-object triggers can be kept low at small bandwidth cost.

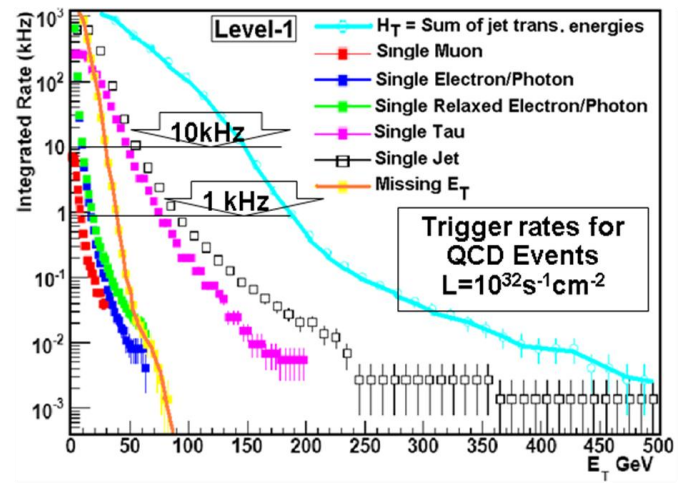

**Figure 2**. L1T rates of single-object triggers as a function of the trigger object transverse momentum threshold.

Based on this information, general guidelines for the bandwidth allocation at L1T can be drawn: (a) muon triggers can be assigned low $P_T$ thresholds at a modest bandwidth cost; (b) electron/photons triggers should be assigned a larger bandwidth, to allow relatively low thresholds, for further and more accurate processing in the HLT (where e.g. tracker information is available); (c) even higher bandwidth should be assigned to energy and jet triggers (also important for energy calibration), to collect data at relatively low $E_T$ thresholds. An example of the resulting bandwidth allocation at L1T is shown in the second column of **Table 3.** If there were only one trigger for a given trigger type, then the bandwidth allocation for that trigger type would determines the trigger threshold. However, one trigger type is in general associated to a set of triggers, all using the same trigger objects (e.g. muons), but applying different threshold values. Then the bandwidth allocated for that trigger stream must be shared among the trigger paths contributing to that stream. The relative prescale

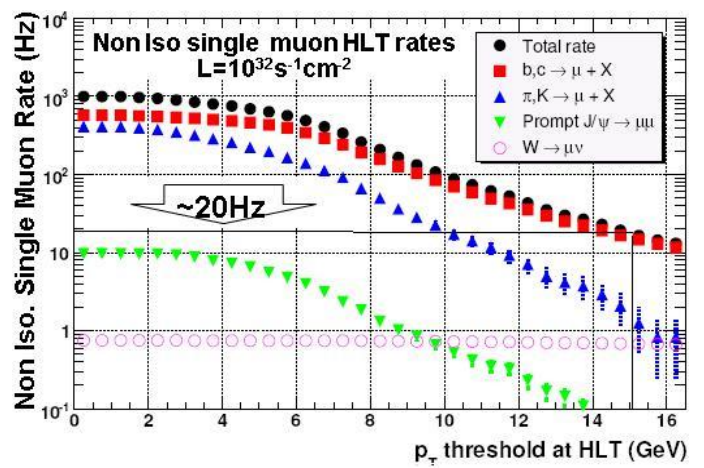

**Figure 3**. HLT rate of the non isolated single muon trigger as a function of the $P_T$ threshold applied at the HLT. Also shown are the contributions from the different processes producing muons

values are set depending on the use of the data samples collected with the lower threshold triggers. A common use is to collect data for trigger efficiency measurements. The relative prescale values are set to collect a suitably large data sample above each threshold, with good event overlap, among samples collected above contiguous thresholds. The lower threshold data will also be used for physics measurements, such as studies of heavy flavour physics down to relatively low muon $P_T$ thresholds, for comparison with measurements at previous colliders.

**4.2 HLT Trigger Rates and Tables**

HLT tables are designed for a maximum HLT accept rate of 150 Hz, which is 50% of the actual initial permanent storage capability of 300 Hz. A reduction factor of two on the allowed output bandwidth is applied for safety.

**Figure 3** shows, as an example, the expected single-muon trigger rate, for $L=10^{32} cm^{-2}s^{-1}$, as a function of the muon $P_T$ threshold for various physics processes. As expected, the low $P_T$ range is dominated by heavy flavor production, while at higher muon $P_T$ values W decays dominate. At this luminosity, an unprescaled muon trigger with a $P_T$ threshold of 15 GeV produces a rate of ~20 Hz. A dimuon trigger with a muon $P_T$ threshold of 3 GeV (the minimum muon $P_T$ detectable in CMS) produces also a rate of ~15 Hz. Single and double muon triggers with low $P_T$ thresholds are important to collect data samples for detector, triggers and physics studies. The low luminosity regime offers a unique opportunity to collect efficiently low $P_T$ muon samples for the study of heavy-flavor and other physics. Thus, at low luminosities, muon triggers are assigned a large portion of the bandwidth.

**Table 3** gives an overview of bandwidth sharing among the different trigger streams at $L=10^{32}s^{-1}cm^{-2}$. Given the relatively low trigger thresholds, affordable at this luminosity, also for unprescaled single object triggers, signal (W/Z, top, Higgs, etc) efficiencies are estimated to be high, between ~70% and ~100% [6], depending on topology.

**4.3 CPU Time Performance**

A key issue for the HLT selection is the CPU power required for the execution of the algorithms in the EF farm. The time performance can be optimized by rejecting events as quickly as possible, using the minimum amount of detector information. For this reason, the basic strategy of a HLT path is to work in "steps" and use partial event reconstruction. The reconstruction of physics objects starts from the corresponding candidates identified by the L1T. Only the parts of the detector pointed to by the L1T information need be considered for further validation of the trigger object. At each step, those parts of each physics object, which can be used for immediate selection, are reconstructed. At the end of each step a set of selection criteria results in the rejection of a significant fraction of the events, while minimizing the CPU usage.

We have measured (on a commercial processor Core 2 5160 Xeon 3.0 GHz), the processing time for running the complete

| Trigger type/stream | Max Allowed Rate L1 (kHz) | HLT (Hz) |
|---|---|---|
| **Muon** (single or double) | 2 | 50 |
| **Electron/photon** (single or double) | 3 | 30 |
| **Single jet or multi-jet and/or Missing Transverse Energy** | 6 | 30 |
| **Tau- and b-jets** | 3 | 20 |
| **Combination of triggering objects** | 3 | 20 |
| **Total output rate** | 17 | 150 |

**Table 3**. Proposed bandwidth allocation to the different trigger types/streams of interest for startup luminosity ($L=10^{32}s^{-1}cm^{-2}$) conditions

HLT table, including the detector data unpacking time, on L1T accepted events from a combination of QCD, heavy flavor and W /Z events, suitably weighted, by their respective expected cross-section, detector acceptance and L1T efficiency. The mean processing time is measured to be 43±6 ms per L1T accepted event.

In the start-up scenario, with DAQ processing capability of 50 kHz of L1 accepted events, an average of ~40 ms per events translates into ~2000 commercial CPUs for the HLT EF farm**.** This was the projected size of the farm from extrapolations back in 2002 at the time of the DAQ/ HLT TDR [4]. We have thus achieved the required CPU time performance of the HLT software.

**5. Conclusions**

The CMS experiment will collect data from the proton-proton collisions delivered by the Large Hadron Collider (LHC) at a centre-of-mass energy of up to 14 TeV, starting operations in Summer 2008. The CMS trigger system is designed to cope with unprecedented luminosities and LHC bunch-crossing rates up to 40 MHz. The unique CMS trigger architecture only employs two trigger levels. The L1T, implemented using custom electronics, inspects events at the full bunch-crossing rate, while selecting up to 100 kHz for further processing. The HLT reduces the 100 kHz input stream to O(100) Hz of events written to permanent storage. The HLT system consists of a large cluster of commercial processors, the Event Filter Farm, running reconstruction and selection algorithms on fully assembled event information. L1 and HLT tables have been developed for startup, low luminosity conditions. A total DAQ readout capability of 50 kHz is assumed at startup. Fast selection and high efficiency is obtained for the physics objects and processes of interest using inclusive selection criteria. The overall CPU requirement is within the system capabilities. In conclusion, the CMS experiment is ready to collect data with high efficiency from the start-up of the LHC operations.